\documentclass[twocolumn,showpacs,preprintnumbers,amsmath,amssymb,superscriptaddress]{revtex4}
\usepackage[cp1251]{inputenc}
\usepackage[ukrainian]{babel}
\usepackage{bm}
\usepackage{epsfig,amssymb,amsmath,bm}

\begin{document}

\title{The hypercomplex equations for fermion interaction description}
\author{K. S. Karplyuk}
\email{karpks@hotmail.com}
 \affiliation{Department of Radiophysics, Taras Shevchenko University, Academic
Glushkov prospect 2, building 5, Kyiv 03122, Ukraine}
\author{O. O. Zhmudskyy}\email{ozhmudsky@physics.ucf.edu}
 \affiliation{Department of Physics, University of Central Florida, 4000 Central Florida Blvd. Orlando, FL, 32816 Phone: (407)-823-4192}
\begin{abstract}

Equations are proposed for the description of the fermion interaction via massive and massless bosons.  
These equations lead to the propagators which maintain theory renormalization.  These equations are also invariant with respect to the 
limited calibration transformations.   Such an approach leads to the possibility of describing fermion interaction via massive bosons and not referring to  
 the spontaneous symmetry violation.   
\end{abstract}


\maketitle
According to the modern point of view the main structural units of matter  are quarks and leptons (fermions with spin $\hbar/2$), which 
interact by boson interchange.  Space-time properties of these fermions are described by Dirac's bispinors.  
These bispinors can be combined into bilinear combinations: scalar $\bar{\psi}\psi$, pseudo-scalar $\bar{\psi}\hat{\iota}\psi$,  
vector $\bar{\psi}\gamma^\alpha\psi$, pseudo-vector $\bar{\psi}\pi^\alpha\psi$, and antisymmetric second-rank tensor   
 $\bar{\psi}\sigma^{\alpha\beta}\psi$.   Here $\gamma^\alpha$ are the Dirac's matrices:       
\begin{equation}
\gamma^\alpha\gamma^\beta+\gamma^\beta\gamma^\alpha=2\eta^{\alpha\beta},
\end{equation}
Here $\eta^{\alpha\beta}=\rm{diag}(1,-1,-1,-1)$, $\hat{\iota}=\gamma^0\gamma^1\gamma^2\gamma^3$, $\pi^\alpha=\gamma^\alpha\hat{\iota}$, $\sigma^{\alpha\beta}=(\gamma^\alpha\gamma^\beta-\gamma^\beta\gamma^\alpha)/2$. 
Some of these bilinear combinations are used as currents which produce the boson fields.  For example, $\bar{\psi}\gamma^\alpha\psi$  
give rise to the electromagnetic field;  $\bar{\psi}\gamma^\alpha\psi$ and $\bar{\psi}i\hat{\iota}\gamma^\alpha\psi$ give rise to the 
$W$- and $Z$ boson fields, which is responsible for the weak interaction.  Other combinations do not have similar use today.   

It is interesting to obtain the general system of fields which can be produced by these bilinear combinations.    
It is convenient to use hypercomplex numbers, based on the Dirac's matrices, in order to obtain this system.  
The system of these hypercomplex numbers contains 16 basic units, represented by matrices.   The system contains the unit matrix $I$ and the 
matrix  $\hat{\iota}$, four matrices $\gamma^\alpha$, four matrices $\pi^\alpha$, and six matrices $\sigma^{\alpha\beta}$.
Hypercomplex numbers of this system are transformed as scalars, pseudo-scalars, vectors, pseudo-vectors and 
antisymmetric second-rank tensor via Lorentz transformations.   In other words, the space of the hypercomplex numbers realizes the reducible representation of the Lorentz group. It is the product of the bispinors representations of $\chi$ and $\bar{\varphi}$.  
 
\begin{gather}
\chi\otimes\bar{\varphi}=\frac{1}{4}\bigl[(\bar{\varphi}I\chi)I-(\bar{\varphi}\hat{\iota}\chi)\hat{\iota}+(\bar{\varphi}\gamma_\alpha\chi)\gamma^\alpha+
(\bar{\varphi}\pi_\alpha\chi)\pi^\alpha-\nonumber\\-\frac{1}{2}(\bar{\varphi}\sigma_{\alpha\beta}\chi)\sigma^{\alpha\beta}\bigr].
\end{gather}
Thus, all bilinear forms $\bar{\psi}\psi$, $\bar{\psi}\hat{\iota}\psi$, $\bar{\psi}\gamma^\alpha\psi$, $\bar{\psi}\pi^\alpha\psi$ and $\bar{\psi}\sigma^{\alpha\beta}\psi$ can be represented by hypercomplex numbers.  In the same way hypercomplex numbers can represent  fields  and potentials produced by these combinations.  

The requirement of the relativistic invariance and the wish to have wave propagators like: 
\begin{equation}
D \sim\frac{1}{k^2-\varkappa^2}
\end{equation}
dictate the choice of the equation, which connects currents and potentials:  
\begin{equation}
(\Box-\varkappa^2)\varPhi=-\left(i\gamma^\nu\partial_\nu-\varkappa\right)\left(i\gamma^\nu\partial_\nu+\varkappa\right)\varPhi=\zeta J.
\end{equation}
Here $\Box=c^{-2}\partial^2/\partial t^2-\triangle$, $\varkappa={mc}/{\hbar}$, $\varPhi$ and $J$ are hypercomplex numbers
\begin{gather}
\varPhi=SI+iA_\alpha\gamma^\alpha+i\varPi_\beta\pi^\beta+\frac{1}{2}\varPhi_{\alpha\beta}\sigma^{\alpha\beta}+\varPsi\hat{\iota},
\end{gather}
\begin{gather}
J=sI+ij_{e\alpha}\gamma^\alpha+ij_{m\beta}\pi^\beta+\frac{1}{2}j_{\alpha\beta}\sigma^{\alpha\beta}+p\hat{\iota}.
\end{gather}
Also $\varPhi_{\alpha\beta}=-\varPhi_{\beta\alpha}$, $j_{\alpha\beta}=-j_{\beta\alpha}$, and the coefficient $\zeta$ is introduced in (4) for 
agreement between unit of currents $J$ and unit of potentials $\varPhi$.  Dirac factorization of the operator  $(\Box-\varkappa^2)$ \cite{d} is used in 
order to write equation  (4).   Let us use two multipliers of this factorization in equation  (4)  in two ways.

At first let us make use the second multiplier in order to connect fields and potentials:  
\begin{equation}
F=-\left(i\gamma^\nu\partial_\nu+\varkappa\right)\varPhi.
\end{equation}
Here $F$ is also a hypercomplex number  
\begin{gather}
F=\epsilon I+iV_\alpha\gamma^\alpha+iU_\beta\pi^\beta+\frac{1}{2}F_{\alpha\beta}\sigma^{\alpha\beta}+\beta\hat{\iota},
\end{gather}
$F_{\alpha\beta}=-F_{\beta\alpha}$. In three dimensional representation equation (7) looks like
\begin{align}
\epsilon&=\frac{1}{c}\frac{\partial A_0}{\partial t}+\nabla\cdot c{\bm A}-\varkappa S,\\
{\bm E}&=-\frac{1}{c}\frac{\partial c{\bm A}}{\partial t}-\nabla A_0+\nabla\times\bm{\varPi}-\varkappa\bm{\theta},\\
c{\bm B}&=\frac{1}{c}\frac{\partial\bm{\varPi}}{\partial t}+\nabla\varPi_0+\nabla\times c{\bm A}-\varkappa\bm{\vartheta},\\
\beta&=\frac{1}{c}\frac{\partial\varPi_0}{\partial t}+\nabla\cdot\bm{\varPi}-\varkappa\varPsi,\\
V_0&=-\frac{1}{c}\frac{\partial S}{\partial t}-\nabla\cdot\bm{\theta}-\varkappa A_0,\\
{\bm V}&=\frac{1}{c}\frac{\partial\bm{\theta}}{\partial t}-\nabla\times\bm{\vartheta}+\nabla S-\varkappa c{\bm A},\\
U_0&=-\frac{1}{c}\frac{\partial{\varPsi}}{\partial t}+\nabla\cdot\bm{\vartheta}-\varkappa\varPi_0,\\
{\bm U}&=-\frac{1}{c}\frac{\partial\bm{\vartheta}}{\partial t}-\nabla\times\bm{\theta}+ \nabla\varPsi-\varkappa\bm{\varPi}.
\end{align}
The usual rules of identification between three-dimensional and four-dimensional coordinates of vectors and tensors leads to:

$\bm{E}=(F_{01},F_{02},F_{03})$, $c\bm{B}=(-F_{23},-F_{31},-F_{12})$, $\bm{V}=(V^1,V^2,V^3)$, $\bm{U}=(U^1,U^2,U^3)$,
$\bm{\theta}=(\varPhi_{01},\varPhi_{02},\varPhi_{03})$, $\bm{\vartheta}=(-\varPhi_{23},-\varPhi_{31},-\varPhi_{12})$, $c\bm{A}=(A^1,A^2,A^3)$, $\bm{\varPi}=(\varPi^1,\varPi^2,\varPi^3)$.  The coefficients $c$ are included in the definition of fields $c\bm{B}$ 
and $c\bm{A}$ in order for the system (9)-(16) to agree with the  $SI$ system of units.

Now let us use the first multiplier in (4) in order to determine field equations and connect them with sources:  
\begin{equation}
\left(i\gamma^\nu\partial_\nu-\varkappa\right)F=\zeta J.
\end{equation}
In a three dimensional representation:
\begin{align}
\frac{1}{c}\frac{\partial {\epsilon}}{\partial t}+\nabla\cdot{\bm E}-\varkappa V_0&=\zeta j_{e0},\\
\frac{1}{c}\frac{\partial {\bm E}}{\partial t}-\nabla\times c{\bm B}+ \nabla\epsilon+\varkappa{\bm V}&=-\zeta{\bm j}_e,\\
-\frac{1}{c}\frac{\partial\beta}{\partial t}+\nabla\cdot c{\bm B}+\varkappa U_0&=-\zeta j_{m0},\\
\frac{1}{c}\frac{\partial c{\bm B}}{\partial t}+\nabla\times {\bm E}-\nabla\beta-\varkappa{\bm U}&=\zeta{\bm j}_m,\\
\frac{1}{c}\frac{\partial V_0}{\partial t}+\nabla\cdot{\bm V}+\varkappa{\epsilon}&=-\zeta s,\\
\frac{1}{c}\frac{\partial{\bm V}}{\partial t}+\nabla V_0-\nabla\times{\bm U}-\varkappa{\bm E}&=\zeta{\bm k},\\
\frac{1}{c}\frac{\partial U_0}{\partial t}+\nabla\cdot{\bm U}+\varkappa{\beta}&=-\zeta p,\\
\frac{1}{c}\frac{\partial{\bm U}}{\partial t}+\nabla U_0+\nabla\times{\bm V}+\varkappa c{\bm B}&=-\zeta{\bm l}.
\end{align}
Here $\bm{k}=(j_{01},j_{02},j_{03})$, $\bm{l}=(-j_{23},-j_{31},-j_{12})$.

Evidently that substitution (9)-(16) into (18)-(25) leads to the equations for potentials (4). In three dimensional representation:
\begin{align}
\square A_0+\varkappa^2 A_0&=\zeta j_{e0},&\square c\bm{A}+\varkappa^2 c\bm{A}&=\zeta\bm{j}_e,\\
\square\varPi_0+\varkappa^2 \varPi_0&=\zeta j_{m0},& \square\bm{\varPi}+\varkappa^2 \bm{\varPi}&=\zeta\bm{j}_m,\\
\square\bm{\theta}+\varkappa^2 \bm{\theta}&=\zeta\bm{k},& \square\bm{\vartheta}+\varkappa^2 \bm{\vartheta}&=\zeta\bm{l},\\
\square S+\varkappa^2S&=\zeta s,& \square\varPsi+\varkappa^2\varPsi&=\zeta p.
\end{align}
We can see that propagators of all fields are in a form (3). This circumstance is important for the quantum theory construction 
because such propagators don't destroy the renormalizability of the theory.  
 
 Equations (18)-(25) can also be obtained in a usual way as Euler-Lagrange equations.   For this we must use the Lagrange density    
\begin{gather}
L=\frac{1}{2\zeta}\Bigl(-\frac{1}{2}F^{\alpha\beta}F_{\alpha\beta}-\epsilon^2+\beta^2+V^\alpha V_\alpha-U^\alpha U_\alpha\Bigr)+\nonumber\\
+\Bigl(sS-p\,\varPsi-j_e^\alpha A_\alpha+j_m^\alpha\varPi_\alpha+\frac{1}{2}j^{\alpha\beta}\varPhi_{\alpha\beta}\Bigr),
\end{gather}
connect fields and potentials with (9)-(16), and treat potentials as independent variables.  

Equations (18)-(25) describe the field system, which can be produced by bilinear combination of the fermion bispinors. 
In the general case it contains a scalar field, a pseudo-scalar field, a vector field, a pseudo-vector field and an antisymmetric second-rank tensor.     
One particular case of this system is the massless electromagnetic field described by Maxwell equations.  Another particular case is the massive 
vector field described by the Proca equation.   

In order to obtain Maxwell's equations from (18)-(25) we must discuss the massless case ($\varkappa=0$), zero all currents except the 
$j_{e\alpha}$, and zero all potentials except the $A_\alpha$.   The only remaining fields are $\bm{E}$, $c\bm{B}$, and $\epsilon$  
and equations for them:  
\begin{align}
\frac{1}{c}\frac{\partial {\epsilon}}{\partial t}+\nabla\cdot{\bm E}&=\zeta j_{e0},\\
\frac{1}{c}\frac{\partial {\bm E}}{\partial t}-\nabla\times c{\bm B}+ \nabla\epsilon&=-\zeta{\bm j}_e,\\
\nabla\cdot c{\bm B}&=0,\\
\frac{1}{c}\frac{\partial c{\bm B}}{\partial t}+\nabla\times {\bm E}&=0.
\end{align}
from (31)-(34) it follows that anomalous field  $\epsilon$ can be produced by non-conserved current only:
\begin{equation}
\Box\,\epsilon=\zeta\Bigl(\frac{1}{c}\frac{\partial j_{e0}}{\partial t}+\mathrm{div}\bm{j}_e\Bigr).
\end{equation}
If the right-hand side of the equation (35) is zero (charge is conserved), we can set up $\epsilon=0$. For zero field $\epsilon$, equations (31)-(34) 
become the Maxwell equations.  In the case $\zeta=\sqrt{\mu_0/\varepsilon_0}$ these equations are written 
in SI system of units.  Thus, Maxwell's equations describe massless bosons born by conserved current only.  On this subject Feynman wrote \cite{feyn}: {\em \lq\lq The laws of physics have no answer to the question: \lq\lq What happens if a charge is suddenly created at this point --- what electromagnetic effects are produced\rq\rq? No answer can be given because our equations say it doesn't happen. If it were to happen, we would need new laws, but we cannot say what they would be.\rq\rq}   Equations (31)-(34) predict what these laws must be.

From (9) we can see that for the massless case, zeroing the field  $\epsilon$ requires the satisfaction of the Lorentz condition:
\begin{equation}
\frac{1}{c}\frac{\partial A_0}{\partial t}+\nabla\cdot c{\bm A}=0.
\end{equation}
In this context we can treat the Lorentz condition as zeroing of the non-conserved currents.   As far as it constrains possible currents it also 
constrains possible potentials.  Thus, it reduces the number of the degrees of freedom of the vector potential from four to three.    

For the massive vector field we must zero all currents except the $j_{e\alpha}$ and also all potentials except the 
 $A_\alpha$.  So, the only remaining fields are  $\bm{E}$, $c\bm{B}$, $\epsilon$, $V_0$, $\bm V$:
\begin{align}
\epsilon&=\frac{1}{c}\frac{\partial A_0}{\partial t}+\nabla\cdot c{\bm A},\\
{\bm E}&=-\frac{1}{c}\frac{\partial c{\bm A}}{\partial t}-\nabla A_0,\\
c{\bm B}&=\nabla\times c{\bm A},\\
V_0&=-\varkappa A_0,\\
{\bm V}&=-\varkappa c{\bm A}.
\end{align}
Equations for these fields become:  
\begin{align}
\frac{1}{c}\frac{\partial {\epsilon}}{\partial t}+\nabla\cdot{\bm E}-\varkappa V_0&=\zeta j_{e0},\\
\frac{1}{c}\frac{\partial {\bm E}}{\partial t}-\nabla\times c{\bm B}+ \nabla\epsilon+\varkappa{\bm V}&=-\zeta{\bm j}_e,\\
\nabla\cdot c{\bm B}&=0,\\
\frac{1}{c}\frac{\partial c{\bm B}}{\partial t}+\nabla\times {\bm E}&=0,\\
\frac{1}{c}\frac{\partial V_0}{\partial t}+\nabla\cdot{\bm V}+\varkappa{\epsilon}&=0,\\
\frac{1}{c}\frac{\partial{\bm V}}{\partial t}+\nabla V_0-\varkappa{\bm E}&=0,\\
\nabla\times{\bm V}+\varkappa c{\bm B}&=0.
\end{align}  
Similar to the massless case, the field $\epsilon$ can be produced by non-conserved currents only:
\begin{equation}
\Box\,\epsilon+\varkappa^2\epsilon=\zeta\Bigl(\frac{1}{c}\frac{\partial j_{e0}}{\partial t}+\mathrm{div}\bm{j}_e\Bigr).
\end{equation}
If charge is conserved, the field $\epsilon$ can be excluded and we obtain field equations in a form:  
\begin{align}
\nabla\cdot{\bm E}-\varkappa V_0&=\zeta j_{e0},\\
\frac{1}{c}\frac{\partial {\bm E}}{\partial t}-\nabla\times c{\bm B}+\varkappa{\bm V}&=-\zeta{\bm j}_e,\\
\nabla\cdot c{\bm B}&=0,\\
\frac{1}{c}\frac{\partial c{\bm B}}{\partial t}+\nabla\times {\bm E}&=0,\\
\frac{1}{c}\frac{\partial V_0}{\partial t}+\nabla\cdot{\bm V}&=0,\\
\frac{1}{c}\frac{\partial{\bm V}}{\partial t}+\nabla V_0-\varkappa{\bm E}&=0,\\
\nabla\times{\bm V}+\varkappa c{\bm B}&=0.
\end{align}
Evidently these equations are equivalent to the Proca equations.  Thus, the Proca equations govern only those vector bosons which are 
the product of the conserved currents only.  As it is known,  unsuccessful attempt to use Proca equation for the non-conserved currents, 
 leads to a propagator different from (3) and destroys renormalizability of the theory  \cite{com},\cite{ok}.      
From the above discussion it is clear that for non-conserved current, equations (42)-(48) must be used instead of the Proca equations.   
Their use guarantees the propagator in a form (3).   

In both cases, Maxwell's equations and the Proca equations, fields are connected with only one potential (the vector potential).  But in the general case 
the same fields (9)-(16) are coupled with several potentials.  For example, as we can see from (10) the \lq\lq electric\rq\rq $\phantom a$ field $\bm{E}$ is connected with the vector potential, the pseudo-vector potential and the tensor potential.  This means that the same fields can be    
produced by different currents.  In other words, fields produced by different currents  interfere with each other.  
Note that electromagnetic field definition via two potentials (vector potential 
and pseudo-vector potential) was used before for the description of the hypothetical magnetic monopole fields \cite{cab}. 

The fields and the potentials introduced above admit constrained calibration transformations.  Should we substitute the potential  $\varPhi$ 
by $\varPhi'$ :    
\begin{equation}
\varPhi'=\varPhi+\left(i\gamma^\alpha\partial_\alpha-\varkappa\right)\varLambda,
\end{equation}
Where $\varLambda$ is the hypercomplex solution of the equation:  
\begin{equation}
(\Box+\varkappa^2)\varLambda=0,
\end{equation}
There is no change of the fields $F$:
\begin{gather}
F=-\left(i\gamma^\alpha\partial_\alpha+\varkappa\right)\varPhi'=\nonumber\\
-\left(i\gamma^\alpha\partial_\alpha+\varkappa\right)\bigl[\varPhi+\left(i\gamma^\alpha\partial_\alpha-\varkappa\right)\varLambda\bigr]=\nonumber\\=
-\left(i\gamma^\alpha\partial_\alpha+\varkappa\right)\varPhi.
\end{gather}
This transformation is constrained because the calibration function $\varLambda$ must satisfy equation (58).  
Let us check how this calibration is working for the two well-know cases, Maxwell's equations and the Proca equations, for the simples case 
of the calibration function $\varLambda$ which contains the scalar part only $\varLambda=\lambda I$.  

In the massless case  ($\varkappa=0$), the previously discussed photon potential $\varPhi=iA_\alpha\gamma^\alpha$ 
receives according to (57) an increment  $i\partial_\alpha\lambda\gamma^\alpha$.  Thus,    
\begin{equation}
A'_\alpha=A_\alpha+\partial_\alpha\lambda,
\end{equation}
Where
\begin{equation}
\Box\lambda=0.
\end{equation}
This is a well-known constrained (special) calibration transformation from electrodynamics.  We can use it for potentials even if  
the Lorentz conditions are already satisfied.  Such a transformation together with the Lorentz  transformation allows us to reduce the number 
of degrees of freedom from four to two.  Such a reduction is necessary for the description of the transverse photons.  

In the massive boson case, the previously discussed potential $\varPhi=iA_\alpha\gamma^\alpha$ is transformed into 
$\varPhi'=-\varkappa\lambda I+i(A_\alpha\gamma^\alpha+\partial_\alpha\lambda)\gamma^\alpha$ according to (57).  Here 
\begin{equation}
\Box\lambda+\varkappa^2\lambda=0.
\end{equation}
It is essential in this case that transformation (57) change not only the the initial vector potential 
\begin{equation}
A'_\alpha=A_\alpha+\partial_\alpha\lambda,
\end{equation}
but also the scalar potential which is initially zero:
\begin{equation}
S'=-\varkappa\lambda.
\end{equation}
Precisely due to this,  fields which are connected with potentials by equalities  (9)-(16) are not changed under this transformation.  
Correspondingly, equations (42)-(48) and the Proca equations (50)-(56) are invariant with respect to this calibration transformation.   
Although they are not invariant with respect to the calibration transformations in traditional sense (if  only $A_\alpha$ is changed).

From the above discussion it follows that the hypercomplex field system (18)-(25) can be used for the description of the interactions  
produced by any fermion currents.  Thus for description of massive and massless bosons which 
mediate interaction between these currents.  
Fermion currents can be either conserved (for example, electric current $\bar{\psi}\gamma^\alpha\psi$), or non-conserved 
(for example, weak interaction  current $\bar{\psi}\gamma^\alpha\gamma^5\psi$).  In either case propagators of the proposed 
fields are in a form (3).   This circumstance guarantees the  renormalizability of the naive  theory which uses the massive bosons.


\end{document}